\documentclass[structabstract]{aa}  
\usepackage{graphicx}
\usepackage{url}
\usepackage{natbib}
\usepackage{lscape}

\usepackage{color}
\usepackage[normalem]{ulem}
\usepackage{txfonts}
\usepackage{xspace}

\newcommand{\ammo}{{\ensuremath{\rm NH_3}}\xspace}

\newcommand{\nh}{{\ensuremath{\rm N_2H^{+}}}\xspace}
\newcommand{\ndp}{{\ensuremath{\rm N_2D^{+}}}\xspace}

\newcommand{\cc}{{\ensuremath{\rm cm^{-3}}}\xspace}

\newcommand{\kms}{{\ensuremath{{\rm km\, s^{-1}}}}\xspace}

\newcommand{\response}[1]{{{ #1}}}

\begin{document}
\title{Spokes cluster: The search for the quiescent gas
\thanks{
This publication is based on data acquired with the Atacama Pathfinder Experiment (APEX) 
under programme 088.F-9322A. 
APEX is a collaboration between the Max-Planck-Institut f\"ur Radioastronomie, 
the European Southern Observatory, and the Onsala Space Observatory.}}
\titlerunning{NGC2264-D APEX}

\author{Jaime E. Pineda\inst{1,2} 
\and 
Paula S. Teixeira\inst{3,4}
}

\authorrunning{J. E. Pineda \& P. S. Teixeira}

\institute{European Southern Observatory (ESO), Garching, Germany
\and
UK ARC Node, Jodrell Bank Centre for Astrophysics, School of Physics and Astronomy, University of Manchester, Manchester, M13 9PL, UK\\
\email{jaime.pineda@manchester.ac.uk}
\and
Universit\"at Wien, Institut f\"ur Astrophysik, T\"urkenschanzstrasse 17, 1180 Vienna, Austria\\
\email{paula.teixeira@univie.ac.at}
\and
Laborat\'orio Associado Instituto D. Luiz-SIM, Universidade de Lisboa, Campo Grande, 1749-016, Lisboa, Portugal.
}

\date{Received ; accepted \today}

\abstract%
{Understanding the role of turbulent and thermal fragmentation is one of the most important current questions of star formation.
To better understand the process of star and cluster formation, we need to study in detail 
the physical structure and properties of the parental molecular cloud. 
In particular, it is important to understand the fragmentation process itself; this may 
be regulated by thermal pressure, magnetic fields, and/or turbulence.
The targeted region, the Spokes cluster, or NGC\,2264\,D,  is a rich protostellar cluster where previous \nh(1--0) 
observations of its dense cores presented linewidths consistent with supersonic turbulence.
However, the fragmentation of the most massive of these cores appears to have a scale length consistent with that of the thermal Jeans length, suggesting that turbulence was not dominant.
}
{
These two results (derived from \nh(1--0) observations and measurements of the spatial separations of the protostars) probe different density regimes. Our aim is to determine if there is subsonic or less-turbulent gas (than previously reported) in the Spokes cluster 
when probing higher densities, which would reconcile both previous observational results.
To study denser gas it is necessary to carry out observations using 
transitions with a higher critical density to directly measure its kinematics.
}
{
We present APEX \nh(3--2) and \ndp(3--2) observations of the NGC2264-D region to measure the 
linewidths and the deuteration fraction of the higher density gas. 
The critical densities of the selected transitions are more than an order of magnitude higher than that of \nh(1--0).
}
{
We find that the \nh(3--2) and \ndp(3--2) emission present significantly narrower linewidths than the emission from \nh(1--0) 
for most cores. 
In two of the spectra, the non-thermal component is close (within 1-$\sigma$) to the sound speed. 
In addition, we find that the three spatially segregated cores, for which no protostar had been confirmed  
show the highest levels of deuteration. 
}
{
These results show that the higher density gas, probed with \nh and \ndp(3--2), reveals more quiescent 
gas in the Spokes cluster than previously reported. More high-angular resolution interferometric observations using high-density tracers are needed to truly assess the kinematics and substructure within NGC2264-D.
} 


\maketitle

\section{Introduction}
\label{intro}

The current paradigm of cluster formation starts with a highly turbulent cloud, in which dense clumps\footnote{Regions that  
may contain substructure and where several stars might form.} are close to virial equilibrium 
\citep[e.g.,][]{McKee_2007-Star_Formation_Review,Bate_2009-Big_HD_Simulation}.
This turbulence will decay in a free-fall time if there is no driving mechanism.
Theoretical studies show that protostellar outflows might be one mechanism 
capable to drive turbulence in the cloud \citep[e.g.][]{Li_2006-Outflow_Turbulence}, 
which agrees with some observational results 
\citep{Palau_2007-Feedback,Fontani_2012-outflow_cavity_creation}. 
However, other studies have shown that outflows might not have enough energy and momentum to sustain turbulence 
\citep{Arce_2010-outflow_Perseus}, while shells driven by young stars might \citep{Arce_2011-Perseus_Shells}.

NGC\,2264, located $\sim$800\,pc away \citep[between 750-913pc,][]{Dahm_2005-TTauri_NGC2264,Baxter_2009-NGC2264_Distance,Naylor_2010-preMS_Older}, is a well-studied cluster 
in which we can study the role of turbulence in star formation. 
Within this cluster there is a rich grouping of protostars, known as NGC\,2264\,D or as the Spokes cluster 
(we use both nomenclatures interchangeably in this paper). 
This region has been observed at (sub)millimeter wavelengths with JCMT/SCUBA and IRAM/30m, where 
several dense cores were identified 
\citep{Ward-Thompson_2000-NGC2264_Dust,Williams_2002-NGC2264_gas,Wolf-Chase_2003-NGC2264_SCUBA,Peretto_2006-NGC2264_MAMBO}. 
 \emph{Spitzer} observations of the region revealed an agglomeration of deeply embedded protostars (Class 0/I sources) identified primarily at 24\,$\mu$m.
\cite{Teixeira_2006-Thermal_Fragmentation}, using a nearest-neighbor analysis, found 
that these protostars have a characteristic projected spacing, 20$\pm$5$\arcsec$, 
which is consistent with the \response{thermal} Jeans length of the region, 
27$\arcsec$ or 0.104\,pc at the distance of 800\,pc 
(calculated using the mean filament density of 3$\times$10$^4$\,cm$^{-3}$ from 
\cite{Williams_2002-NGC2264_gas}, and temperature of 17\,K
from \cite{Ward-Thompson_2000-NGC2264_Dust}). 
Since the protostars are embedded in dense material, and the virial mass of the system is lower than the total gas and dust mass 
\citep{Williams_2002-NGC2264_gas,Wolf-Chase_2003-NGC2264_SCUBA,Peretto_2006-NGC2264_MAMBO}, 
the protostars are bound to the filaments and are very likely still tracing the primordial substructure of the cluster. 
The measured regular spacing (consistent with the \response{thermal} Jeans length) of the protostars is thus interpreted as a fossil signature of thermal fragmentation of the parental filament \citep{Teixeira_2006-Thermal_Fragmentation}.

\cite{Peretto_2006-NGC2264_MAMBO} also observed the 15 cores in the Spokes cluster 
using dust continuum emission. 
These dense cores present a mean ${\rm H_2}$ density ranging between 
$10^5$ and $10^6$\,\cc, and masses of 1.8 to 17.3\,M$_\odot$ 
\citep{Wolf-Chase_2003-NGC2264_SCUBA,Peretto_2006-NGC2264_MAMBO}. 
Direct comparison with the \emph{Spitzer} 24\,$\mu$m image shows that the cores identified by \cite{Peretto_2006-NGC2264_MAMBO} have
24\,$\mu$m \emph{Spitzer} counterparts
(i.e., they harbor protostars) except for DMM-8 and DMM-11. 
\cite{Peretto_2006-NGC2264_MAMBO} also observed these 15 dense cores in 
\nh(1--0), which traces high-density gas (2$\times10^{5}$\,\cc), and found only lines with supersonic 
velocity dispersions (Mach $\sim$3), which is at odds with the interpretation of \response{thermal} Jeans fragmentation. 
Follow-up observations by \cite{Teixeira_2007-NGC2264_SMA} using the  Submillimeter Array \citep{Ho_2004-SMA}
of one of the cores, DMM-1, 
showed that it had fragmented into seven compact 1.3\,mm sources, of masses ranging between 0.4\,M$_\odot$ and 1.2\,M$_\odot$. The mean separation of these sources within DMM-1 is consistent with the \response{thermal} Jeans length of the core, which indicates that the core underwent thermal fragmentation.

The apparent disagreement between the observation of supersonic turbulence in the dense gas and the thermal fragmentation, 
suggested by the spatial separation of the young stellar objects (YSOs), can be tested. 
If thermal fragmentation indeed took place in the star formation process in this cluster, 
low levels of turbulence should still be seen using a higher-density tracer. 
The Spokes cluster contains high-density gas over an extended region, 
which could explain the \nh(1--0) observations by \cite{Peretto_2006-NGC2264_MAMBO}, 
which show supersonic turbulence in a dense gas tracer. These 
observations might not be tracing the densest gas in the star-forming cores, but instead it traces the dense environment containing the entire region. 
This interpretation can be readily tested with new observations using an 
even higher-density tracer at similar angular resolutions:  \nh(3--2) or \ndp(3--2). 
\response{Moreover, the [\ndp/\nh] abundance ratio is enhanced in the densest and coldest regions of cores 
\citep{Caselli_2002-Deuteration_conf}, and it decreased after the protostar was formed \citep{Emprechtinger_2009-deuteration_class0}. 
Therefore, the \ndp(3--2) observations trace more pristine conditions than \nh(3--2).}

A decrease in the velocity dispersion is expected when going to higher-density 
tracers, as shown by \cite{Myers_1983-subsonic_turbulence}, \cite{Goodman_1998-coherence}, and 
\cite{Fuller_1992-dv_r} in regions with lower average densities. 
Moreover, \cite{Pineda_2010-transition_coherence} recently presented observations of 
dense gas as traced by \ammo(1,1) where a sharp transition from supersonic to subsonic 
turbulence is seen in the velocity dispersion. 
This result suggests that to form low-mass stars, the turbulence needs to be dissipated 
before the star can be formed. 
It is possible that in more massive and denser environments this transition would occur at 
higher densities, and therefore higher-density tracers need to be used.
\citet{Foster_2009-GBT} showed, from their  observations of the dense cores in the Perseus molecular cloud, that 
the cores have subsonic \ammo(1,1) linewidths regardless of their stellar content or cluster environment. 
Moreover, even though the effect of the molecular outflows is important in the dynamics 
of the low-density gas, the dense gas is only locally affected 
\citep{Pineda_2010-transition_coherence,Pineda_2013-GBT_Maps,Pineda_2011-EVLA_Filament}.

Therefore, by observing the Spokes cluster in a denser-gas tracer 
than \nh(1--0), we can determine whether the emission arising from denser regions is 
less turbulent. This could provide a test of the thermal-fragmentation  hypothesis  of the quiescent gas.

\section{Data}

The observations of \nh and \ndp(3--2) molecular lines were carried out on 
5 September and 4, 8-9, and 11-13 November 2011 with APEX. 
Every dense core in the Spokes cluster identified by 
\cite{Peretto_2006-NGC2264_MAMBO} that presented a good detection of the 
\nh(1--0) line (14 cores)
was observed in \nh(3--2) using the APEX-2 receiver 
of the Swedish Heterodyne Facility Instrument 
\citep[SHeFI;][]{Vassilev_2008-APEX_SHeFI}. 
The cores with the narrowest \nh(1--0) lines (eight cores) were also observed in \ndp(3--2) using the APEX-1 receiver 
of the SHeFI  \citep{Vassilev_2008-APEX_SHeFI}.

\begin{figure}[!h]
\centering
\includegraphics[width=0.47\textwidth]{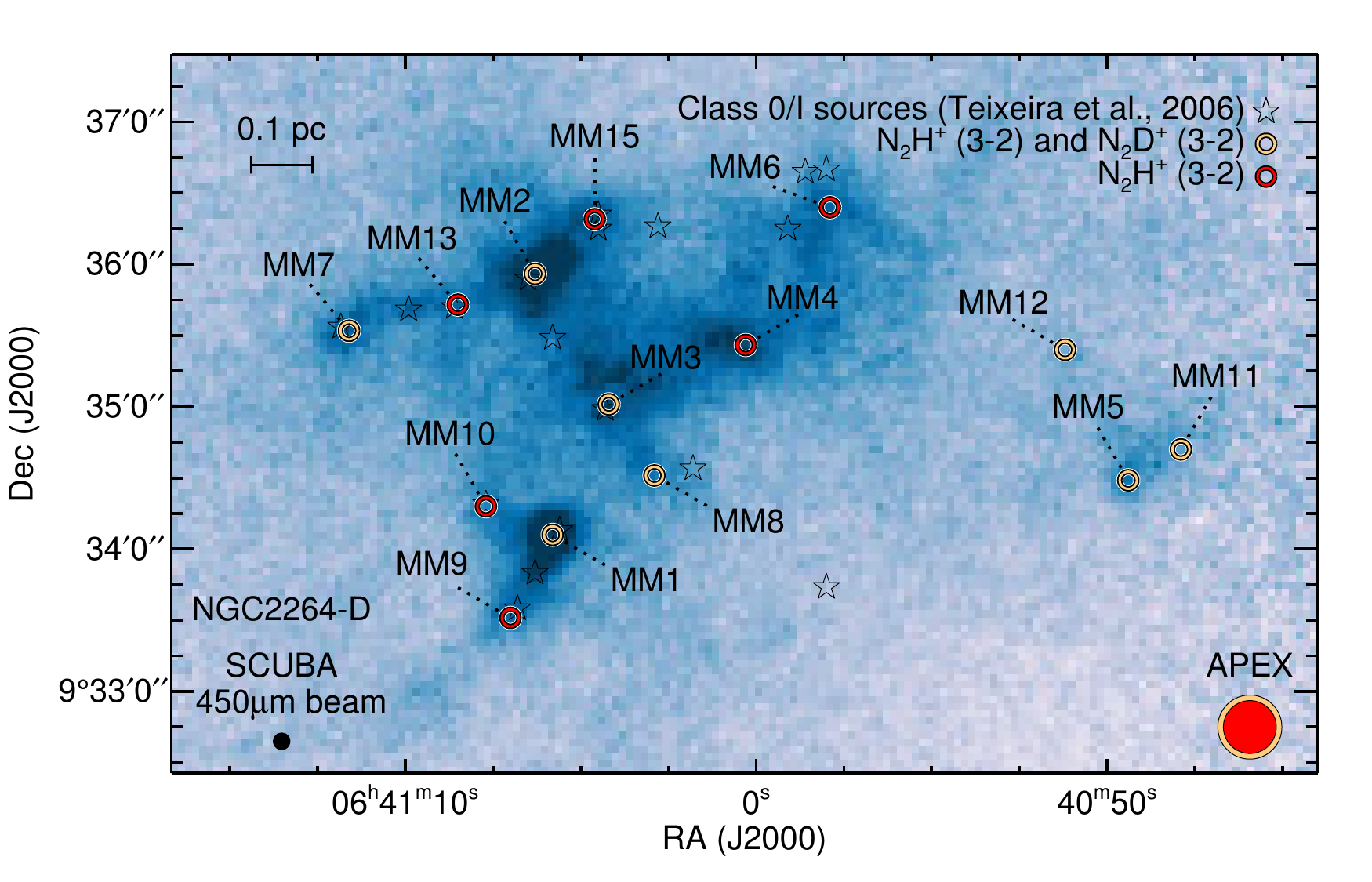}
\caption{SCUBA 450$\mu$m dust-continuum map for the Spokes cluster 
\citep{Wolf-Chase_2003-NGC2264_SCUBA}. 
The observed dense cores identified in \cite{Peretto_2006-NGC2264_MAMBO} are shown by the circles at their centers
(color-coded according to which molecular line tracers were used in their observation). For comparison, the APEX beams corresponding to our observations are also shown in the lower right corner (following the same color-code convention).
\label{fig-map}}
\end{figure}

Figure~\ref{fig-map} shows a JCMT/SCUBA 450\,$\mu$m map of the Spokes cluster,  including the locations of 
the previously identified protostars and our APEX observations.
A summary of the sources  is listed in Table~\ref{obs-sum}.
The observed transitions, their spectroscopic properties, and observational parameters
are listed in Table~\ref{table-telescope}. We use the following nomenclature for source identification: 
MM??, where `??' is the source number (e.g., source D-MM1 of \citet{Peretto_2006-NGC2264_MAMBO} is referred to as MM01 hereafter).

The observations used position-switching, with the off-position at 
($\alpha_{J2000}$=$06^{\rm h}40^{\rm m}52.57^{\rm s}$, $\delta_{J2000}$=$09\degr32\arcmin36.5\arcsec$). 
They were calibrated by regularly measuring the sky brightness and cold/hot loads in the cabin.
All observations used the  eXtended bandwidth Fast Fourier Transform Spectrometer (XFFTS), 
which provides two 2.5\,GHz units, with 32,768 channels each unit and 1\,GHz overlap. 
This configuration provides an spectral resolution of $0.082$\kms and $0.099$\kms for \nh and \ndp(3--2), respectively.

All spectra were reduced using \verb+CLASS90+\footnote{\url{http://www.iram.fr/IRAMFR/GILDAS}}. 
Each spectrum had a first- or second-order polynomial fitted to remove the baseline 
before averaging all scans.

\begin{table*}[!h]
\centering
\caption{List of observed sources\label{obs-sum}}
\begin{tabular}{lcccccc}
\hline\hline
&&&&\multicolumn{2}{c}{On-source time}&\\
Source &	R.A.&	Decl. &	$\sigma_{v}$\tablefootmark{a}	&  \nh(3--2)  & \ndp(3--2) & \# YSOs\\
 & (hh:mm:ss) & (dd:mm:ss) & 	(\kms)	& (min) & (min) &  within APEX beams\tablefootmark{b} \\
\hline
MM01	& 06:41:05.8	& +09:34:06	&0.59&   4.0 & 23.0 & 1 (7)\tablefootmark{c}\\
MM02	& 06:41:06.3	& +09:35:56	&0.51&   2.3 & 28.0 & 1\\
MM03	& 06:41:04.2	& +09:35:01 	&0.26&   3.3 & 24.0 &1\\
MM04	& 06:41:00.3	& +09:35:26 	&0.81&   3.3 & \ldots &0 (1)\tablefootmark{d}\\
MM05	& 06:40:49.4	& +09:34:29	&0.42& 13.4 & 24.0 &0 (1)\tablefootmark{d}\\
MM06	& 06:40:57.9	& +09:36:24 	&0.55&   3.3 & \ldots &0 (1)\tablefootmark{d}\\
MM07	& 06:41:11.6	& +09:35:32 	&0.59&   3.3 & 17.0 &1\\
MM08	& 06:41:02.9	& +09:34:31 	&0.34&   3.3 & 31.0 &0\\
MM09	& 06:41:07.0	& +09:33:31 	&0.55&   3.3 & \ldots &1\\
MM10	& 06:41:07.7	& +09:34:18 	&0.72&   3.3 & \ldots &1\\
MM11	& 06:40:47.9	& +09:34:42 	&0.30& 13.7 & 33.0 &0\\
MM12	& 06:40:51.2	& +09:35:24 	&0.38&   8.0 & 79.0 &0 (1)\tablefootmark{d}\\
MM13	& 06:41:08.5	& +09:35:43 	&0.89&   6.3 & \ldots &1\\
MM15	& 06:41:04.6	& +09:36:19	&0.64& 13.7 & \ldots &2\\
\hline
\end{tabular}
\tablefoot{
\tablefoottext{a}{Velocity dispersion from \cite{Peretto_2006-NGC2264_MAMBO} derived using \nh(1--0) fits.}\\
\tablefoottext{b}{Number of protostars located within the APEX HPBW (see Table \ref{table-telescope}) \citep[][and references therein]{Peretto_2006-NGC2264_MAMBO}, \citep{Teixeira_2006-Thermal_Fragmentation}.}\\
\tablefoottext{c}{Targeted SMA observations of this core revealed additional compact millimeter sources undetected in the infrared \citep{Teixeira_2007-NGC2264_SMA}.}\\
\tablefoottext{d}{The value in parenthesis corresponds to the number of faint 24\,$\mu$m sources that have no other wavelength counterpart  (and therefore we cannot confirm their protostellar nature).}
}
\end{table*}

\begin{table*}[!h]
\centering
\caption{Observed transitions and observational parameters\label{table-telescope}}
\begin{tabular}{lccccccc}
\hline\hline
Transition & Frequency & HPBW & \multicolumn{2}{c}{channel spacing} & $\eta_{MB}$ & B & $\mu$\\
                      & (GHz) & (\arcsec) &  (kHz) & (\kms) & & (MHz)& (D)\\
\hline
\ndp(3--2)  & 231.3219119 & 27.0 & 76.29 & 0.0989 & 0.75 & 38554.719 & 3.4\\
\nh(3--2)    & 279.511862 & 22.3 & 76.29 & 0.0818  & 0.74 & 46586.867 & 3.4\\
\hline
\end{tabular}
\tablefoot{
$\eta_{MB}$ is the main-beam efficiency of the APEX telescope, B is the rotational constant of the molecule, and $\mu$ is the dipole moment of the molecule. 
}
\end{table*}

\section{Results}

\begin{figure*}[!h]
\includegraphics[width=\textwidth]{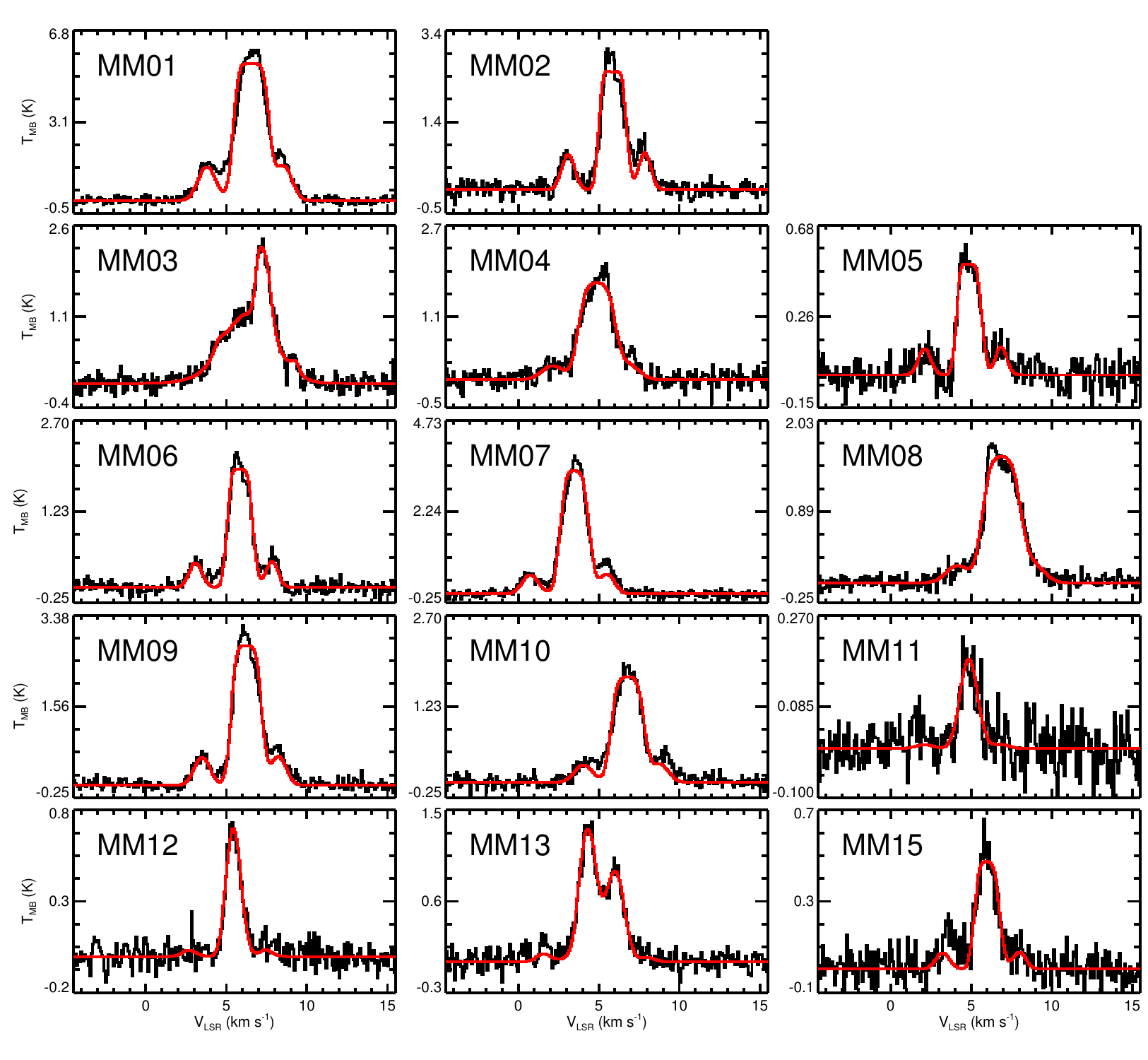}
\caption{Observed \nh(3--2) spectra toward the sources. The red curve is the best-fit model to the observations, whose fit parameters are shown in Table \ref{n2h-n2d-fit}.
\label{fig-n2h+-spectra}}
\end{figure*}

\begin{figure*}[!h]
\includegraphics[width=\textwidth]{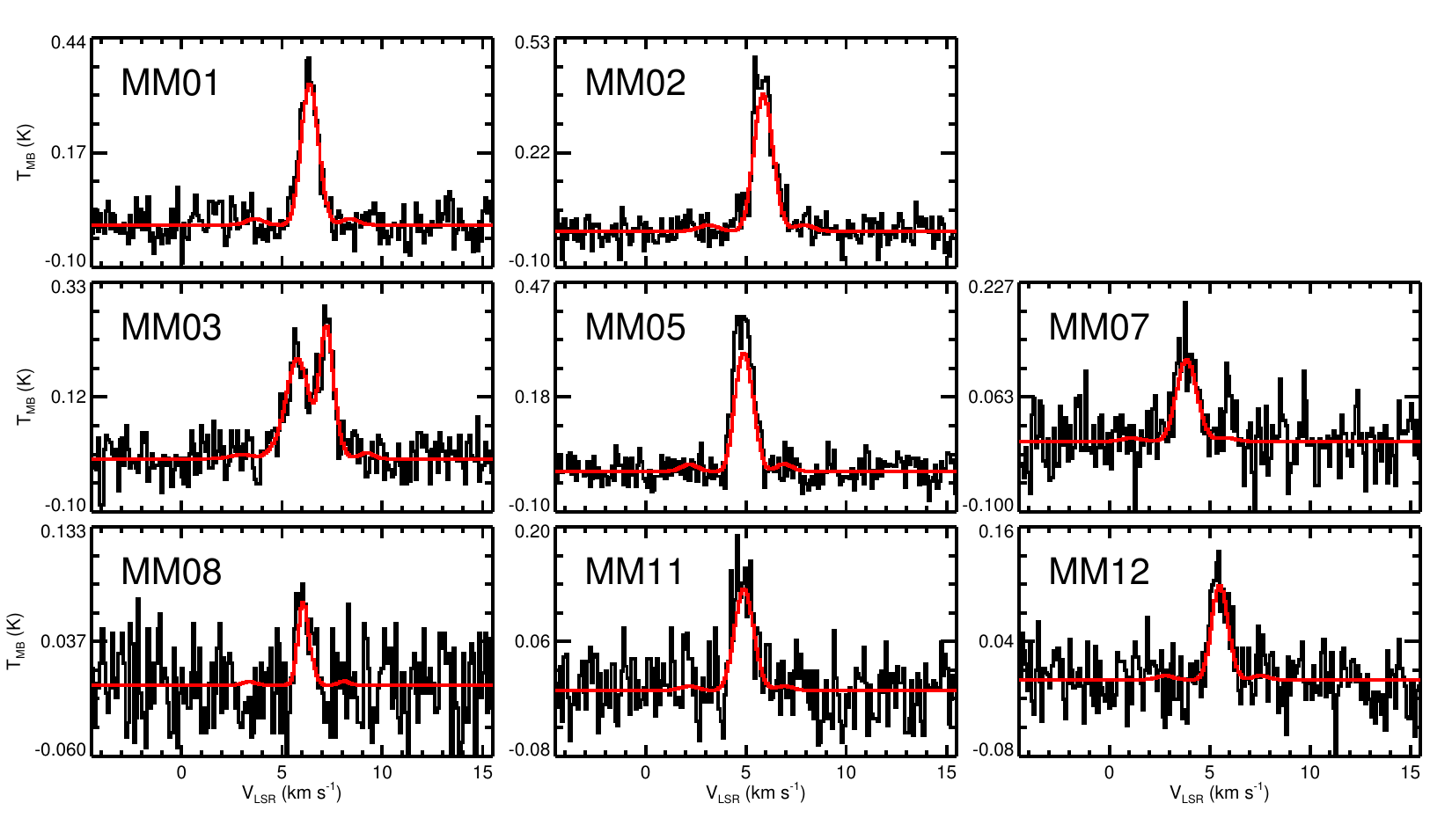}
\caption{Observed \ndp(3--2) spectra toward the sources. The red curve is the best-fit model to the observations, whose fit parameters are shown in Table \ref{n2h-n2d-fit}.
\label{fig-n2d+-spectra}}
\end{figure*}

We detected \nh (3--2) and \ndp(3--2) emission toward all 14 and 8 observed cores respectively.
The \nh(3--2) and \ndp(3--2) spectra are shown in Figures~\ref{fig-n2h+-spectra} and 
\ref{fig-n2d+-spectra}, respectively. In addition, we overplot on these figures the best-fit profiles to the lines.

The \nh and \ndp (3--2) lines were fitted in IDL\footnote{Interactive Data Language} using a 
forward-fitting model that takes into account all hyperfine components, with the 
minimization performed using the \verb+mpfitfun+ program \citep{IDL_MPFIT}. 
For a given centroid velocity ($v_{LSR}$), velocity dispersion ($\sigma_v$), 
and total optical depth ($\tau_{main}$) we computed the optical depth at each velocity as
\begin{equation}
\tau = \tau_{main} \Sigma_{i} \,w_i \exp{-(v-v_i-v_{LSR})^2/(2\sigma_v^2)}~,
\end{equation}
where $w_i$ and $v_i$ are the relative weight and offset velocity for each hyperfine component.  
The model spectrum is then generated as 
\begin{equation}
T_{MB} = \eta_f\, [J(T_{ex}) - J(T_{bg})] \left[1-e^{-\tau}\right]
\end{equation}
where $T_{MB}$ is the main-beam temperature, 
$\eta_f$ is the filling fraction of the emission in the beam, 
$T_{ex}$ is the excitation temperature, 
$T_{bg}$ is the background brightness temperature (2.73~K), and 
\[
J(T)= \frac{h\nu}{k} ~ \frac{1}{\exp{(h\nu/k T)}-1}~. 
\]
\response{Since the extent of the emission is unknown and it is unclear if it is more or less extended than the 
continuum emission, we assumed a filling fraction of unity, $\eta_f=1$.}
Therefore, $T_{ex}$ is obtained from the fit when $\tau_{main}$ is well constrained. 
We also compared the results of fitting the spectra using the HFS mode within \verb+CLASS90+, 
where we obtained the same results within the fitted parameter uncertainties. 
This procedure usually returns well-constrained parameters, but when the total optical depth is 
poorly constrained ($\tau < 3\sigma_{\tau}$), we kept the excitation temperature as a fixed parameter. 
For the \nh (3--2) line, the excitation temperature was set to the average value obtained from 
all other spectra, i.e., $6.7$\,K, because this average excitation temperature provides the 
typical excitation conditions in the region. 
For \ndp (3--2), \response{the fit is only poorly constrained and therefore} 
the excitation temperature was set to the value obtained from the \nh (3--2) fit, 
because both transitions are optically thin and have similar critical densities. 
This procedure is similar to that followed in \cite{Crapsi_2005-N2H_N2D}.

For most of the cores the fits are good, but there are 
two cores (MM03 and MM13) that present two distinct components along the line-of-sight. 
In this case two independent lines were fitted simultaneously, and each of the two components was labeled 
``a'' or ``b''.
There are two other cores, MM08 and MM04, which show a clear asymmetry in the line profile that suggestive of 
an additional component along their line-of-sight. Unfortunately, we were unable to separate the blended components.
Table~\ref{n2h-n2d-fit} shows the results of the hyperfine fitting procedure to the
\nh(3--2) and \ndp(3--2) lines of all sources.

\begin{table*}[!h]
\caption{Best-fit line parameters for sources in NGC2264-D}
\label{n2h-n2d-fit}
\centering
\begin{tabular}{lcccccccccc}
\hline\hline
&\multicolumn{5}{c}{\nh(3--2)}&&\multicolumn{4}{c}{\ndp(3--2)} \\\cline{2-6}\cline{8-11}
Source\tablefootmark{a} &  $T_{\rm ex}$  & $V_{LSR}$ &   $\sigma_v$   &  $\tau_{main}$ & W &&
                            $V_{LSR}$ &   $\sigma_v$   &  $\tau_{main}$ & W \\
  &  (K) &   (\kms) &   (\kms) & & (K\,\kms) &
            &   (\kms) &   (\kms) & & (K\,\kms) \\
\hline
MM01  &  10.90$\pm$0.03   &  6.405$\pm$0.004  & 0.495$\pm$0.003 &   8.0$\pm$0.2 &16.07$\pm$0.08&
       &   6.26$\pm$0.02   &  0.38$\pm$0.02  & $<$0.07  &  0.43$\pm$0.03\\
MM02  &   7.37$\pm$0.05   &  5.713$\pm$0.009  & 0.339$\pm$0.007 &  10.5$\pm$0.6 & 5.9$\pm$0.1&
       &   5.75$\pm$0.02   &  0.43$\pm$0.02  & $<$0.19  &  0.57$\pm$0.02\\
MM03a &   5.6$\pm$0.2   &  7.15$\pm$0.02  & 0.22$\pm$0.03 &   4$\pm$1 & 5.89$\pm$0.09&
       &   7.11$\pm$0.03   &  0.28$\pm$0.03  & $<$0.27  &  0.52$\pm$0.03\\
MM03b & 6.7 &  6.25$\pm$0.05  & 1.33$\pm$0.04 & $<$0.90  &  \ldots\tablefootmark{b}&
       &   5.65$\pm$0.05   &  0.54$\pm$0.06  & $<$0.11  &  \ldots\tablefootmark{b}\\
MM04  &   6.34$\pm$0.07   &  4.75$\pm$0.02  & 0.61$\pm$0.02 &   4.2$\pm$0.5 & 4.8$\pm$0.1&
       & \ldots & \ldots &\ldots &\ldots \\
MM05  &   4.25$\pm$0.03   &  4.72$\pm$0.02 & 0.33$\pm$0.02 &   8$\pm$1 & 1.04$\pm$0.04&
       &   4.79$\pm$0.02   &  0.36$\pm$0.02  & $<$1.04  &  0.50$\pm$0.03\\
MM06  &   6.58$\pm$0.03   &  5.694$\pm$0.007  & 0.349$\pm$0.007 &   6.9$\pm$0.4 & 3.96$\pm$0.06&
       & \ldots & \ldots &\ldots &\ldots \\
MM07  &   8.53$\pm$0.03   &  3.366$\pm$0.005  & 0.452$\pm$0.005 &   4.7$\pm$0.2 & 7.53$\pm$0.06&
       &   3.72$\pm$0.07   &  0.47$\pm$0.07  & $<$0.04  &  0.19$\pm$0.03\\
MM08  &   6.17$\pm$0.05   &  6.75$\pm$0.01  & 0.65$\pm$0.02 &   3.9$\pm$0.4 & 4.59$\pm$0.06&
       &   5.97$\pm$0.06   &  0.23$\pm$0.07  & $<$0.06  &  0.07$\pm$0.03\\
MM09  &   7.74$\pm$0.03   &  6.103$\pm$0.006  & 0.436$\pm$0.005 &   6.4$\pm$0.2 & 6.76$\pm$0.06&
       & \ldots & \ldots &\ldots &\ldots \\
MM10  &   6.32$\pm$0.04   &  6.66$\pm$0.01  & 0.51$\pm$0.01 &   5.0$\pm$0.4 & 4.38$\pm$0.06&
       & \ldots & \ldots &\ldots &\ldots \\
MM11  & 6.7 &  4.72$\pm$0.05  & 0.50$\pm$0.05 & $<$0.11  &  0.27$\pm$0.03&
       &   4.78$\pm$0.05   &  0.43$\pm$0.06  & $<$0.07  &  0.19$\pm$0.03\\
MM12  & 6.7 &  5.31$\pm$0.02  & 0.39$\pm$0.02 & $<$0.52  &  0.89$\pm$0.04&
       &   5.38$\pm$0.05   &  0.36$\pm$0.05  & $<$0.06  &  0.13$\pm$0.02\\
MM13a & 6.7 &  4.20$\pm$0.02  & 0.42$\pm$0.02 & $<$1.20  &  3.14$\pm$0.07&
       & \ldots & \ldots &\ldots &\ldots \\
MM13b & 6.7 &  5.84$\pm$0.03  & 0.49$\pm$0.03 & $<$0.61  &  \ldots\tablefootmark{b}&
       & \ldots & \ldots &\ldots &\ldots \\
MM15  &   4.18$\pm$0.05   &  5.85$\pm$0.02  & 0.38$\pm$0.03 &   5$\pm$1 & 0.89$\pm$0.05&
       & \ldots & \ldots &\ldots &\ldots \\
\hline
\end{tabular}
\tablefoot{
\tablefoottext{a}{If the spectrum shows two components, they were both fitted, listed in different rows, and 
labeled as components a and b.}\\
\tablefoottext{b}{We were unable to derive W for each of the two components -- the value listed for component a corresponds to the integrated W for both components.}
}
\end{table*}

The column densities were calculated using the constant excitation temperature approximation 
and that the beam filling factor is $1$ for both lines, see 
equations~(\ref{eq-op-thick}) and (\ref{eq-op-thin}) from Appendix~\ref{sec-app-a}, which are based on 
 \cite{Caselli_2002-Ionization_L1544}.

\begin{table}[!h]
\caption{Column densities
\label{table-col-dens}}
\centering
\begin{tabular}{lccc}
\hline\hline
Source & $N(\nh)$ & $N(\ndp)$ & $N(\ndp)/N(\nh)$ \\
 &  $(10^{12} {\rm cm^{-2}})$  & $(10^{11} {\rm cm^{-2}})$ & \\
\hline
MM01  &  58$\pm$1   & 
 4.4$\pm$0.3  & 0.0076$\pm$0.0006\\
MM02  &  56$\pm$4   & 
11.2$\pm$0.5  & 0.020$\pm$0.002\\
MM03$^{\dagger}$ &  19.5$\pm$0.3   & 
12.7$\pm$0.7  & 0.065$\pm$0.004\\
MM04  &  45$\pm$6   & 
  \ldots & \ldots \\
MM05  &  86$\pm$14  & 
65$\pm$3  & 0.08$\pm$0.01\\
MM06  &  41$\pm$3   & 
  \ldots & \ldots \\
MM07  &  31$\pm$1   & 
 2.7$\pm$0.4  & 0.009$\pm$0.002\\
MM08  &  46$\pm$5   & 
2.0$\pm$0.9  & 0.004$\pm$0.002\\
MM09  &  42$\pm$2   & 
  \ldots & \ldots \\
MM10  &  45$\pm$3   & 
  \ldots & \ldots \\
MM11  &   0.9$\pm$0.1   & 
 4.6$\pm$0.7  & 0.5$\pm$0.1\\
MM12  &   3.0$\pm$0.1   & 
 3.2$\pm$0.5  & 0.11$\pm$0.02\\
MM13$^{\dagger}$ &   10.4$\pm$0.2   & 
  \ldots & \ldots \\
MM15  &  60$\pm$15  & 
  \ldots & \ldots \\ 
  \hline
\end{tabular}
\tablefoot{
\tablefoottext{\dagger}{The spectra of sources NGC2264D-MM03 and -MM13 show two components, but since 
the optically thin model had to be used for at least one of the components, we used the optically thin column density estimate. 
}
}
\end{table}

\begin{figure*}
\sidecaption
\includegraphics[width=12cm]{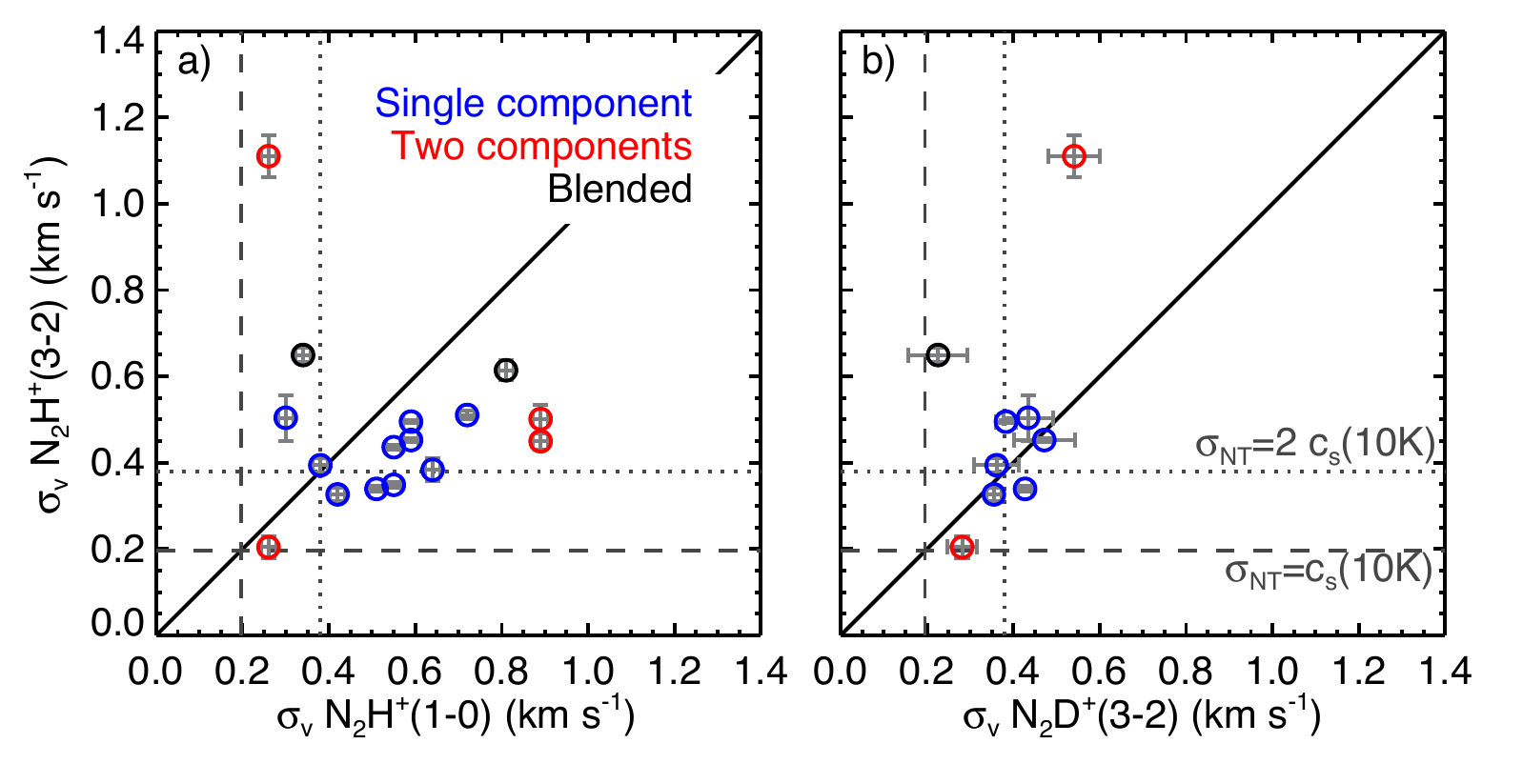}
\caption{Comparison of the velocity dispersion of different molecular lines. 
Cores listed as showing two components (see Table~\ref{n2h-n2d-fit}) are shown as red symbols, 
cores that appear to have blended components are shown in black,
 and the rest are shown as blue symbols. 
Dashed and dotted lines show the expected velocity dispersion for the cases where the nonthermal velocity dispersion, 
$\sigma_{NT}$, equals $c_{s}(10\,{\rm K})$ or $2\,c_{s}(10\,{\rm K})$, where $c_{s}(10\,{\rm K})$ is the sound speed of the 
average molecule at 10~K (for a mean molecular weight of $\mu_m=2.33$).
\label{fig-plot-sigma}}
\end{figure*}

Figure~\ref{fig-plot-sigma} compares the velocity dispersions obtained from the different molecular lines that we observed 
and the \nh (1--0) velocity dispersions reported in \citet{Peretto_2006-NGC2264_MAMBO}. 
The velocity dispersion, $\sigma_v$, has a thermal, $\sigma_\mathrm{T}$,  and a nonthermal, 
$\sigma_\mathrm{NT}$, component and is given as 
$\sigma_v = \sqrt{\sigma_\mathrm{T}^2 + \sigma_\mathrm{NT}^2}$.
Each panel also indicates, for comparison purposes, expected velocity dispersions whose nonthermal 
components are $\sigma_{NT}=c_{s}(10\,{\rm K})$ (dashed lines)
and $\sigma_{NT}=2\,c_{s}(10\,{\rm K})$ (dotted lines). 
Figure~\ref{fig-plot-sigma}-a shows the comparison of the velocity dispersion obtained from the two different transitions
\nh(1--0) and (3--2); the plot clearly shows \emph{that there is a trend for the velocity dispersion to be narrower for the higher-J transition}. Figure~\ref{fig-plot-sigma}-b shows the comparison of the velocity dispersions of the \nh and \ndp (3--2) molecular lines. 
From this plot we see that the velocity dispersions of these two high-density tracers agree well.

\section{Discussion and conclusion}

As stated in section \S\,\ref{intro}, we aimed to test whether the previous \nh(1--0) observations of the Spokes cluster 
were in fact probing the larger extended emission and not the densest gas in the star-forming cores. 
If this scenario is correct, we would expect the linewidths of higher-J transition emission, tracing higher-density gas, 
to be narrower. 
The results presented in the previous section indicate that the denser gas does indeed present narrower linewidths. 
For \nh(1--0) lines, as previously discussed in \cite{Peretto_2006-NGC2264_MAMBO}, 
the nonthermal velocity dispersions are supersonic and in most of them the $\sigma_{NT}>2\,c_{s}(10~{\rm K})$.
However, in the higher-density gas, traced by \nh(3--2), the nonthermal velocity dispersion is lower than that of \nh(1--0).
In fact, the \nh(3--2) velocity dispersions are $\approx$70\% of the value derived from \nh(1--0), and more than half of the 
total sample shows $\sigma_{NT}\le2\,c_{s}(10~{\rm K})$. 
There are, however, three outlier data points where the velocity dispersion derived from 
\nh(3--2) is higher than those from \nh(1--0). 
Close inspection reveals that 
a) one of these outliers corresponds to a poorly constrained second component (MM03b), 
b) another outlier presents asymmetries in the line profile that clearly suggests the presence of a second 
unresolved component along the line-of-sight (MM08), and finally, 
c) the remaining outlier data point is obtained from the spectrum with the lowest signal-to-noise ratio with a 
large associated uncertainty (MM11). 
Finally, the velocity dispersions obtained with \ndp(3--2)  either agree with those obtained with \nh(3--2) or are lower. 
It is important to note there are two cores in our sample  that show velocity dispersions consistent 
(within the 1-$\sigma$ uncertainties) with a sonic nonthermal component: 
MM03a in \nh(3--2) and MM08 in \ndp (3--2).

These results thus support the thermal-fragmentation scenario presented in \cite{Teixeira_2006-Thermal_Fragmentation}. 
Although the nonthermal velocity dispersions here reported are still broad, they are much narrower than those found 
previously using \nh(1--0) single-dish observations \citep{Peretto_2006-NGC2264_MAMBO}. 
It also is important to note that these single-dish observations have a coarse angular resolution 
($\approx17\,840\,{\rm AU}= 0.086\,{\rm pc}$ at 800\,pc), and the substructure or several components along the line-of-sight 
could increase the observed linewidth.  
It is clear that more high-angular resolution interferometric observations using high-density tracers are 
needed to truly assess the kinematics and substructure within NGC2264-D.

Deuteration can be used as a proxy for the chemical evolution of a core 
\citep{Crapsi_2005-N2H_N2D,Emprechtinger_2009-deuteration_class0,Fontani_2011-Deuteration_MSF}. 
For prestellar cores, the deuteration fraction increases as these cores evolve, reaching a maximum at the onset of star formation; the deuteration fraction of the core then begins to decrease as the protostar evolves. For this reason, a high-deuteration fraction in a protostellar core is a particular indicator of youth of the protostar, and a high-deuteration fraction of a prestellar core is an indicator that the core may soon collapse to form a protostar. 
Table~\ref{table-col-dens} shows that there are four cores (MM03, MM05, MM11, and MM12) with a 
deuteration fraction at least a factor of $3$ higher than the rest of the sample, where the largest difference is 
a factor of 125. 
Three of these cores are spatially separated 
from the main cluster (see Figure~\ref{fig-map}) and are embedded in a less dense region. 
Furthermore, Table~\ref{obs-sum} shows that no protostar is identified in MM11, and both MM05 and MM12 have faint 24\,$\mu$m sources that have no counterparts at other infrared wavelengths, meaning that these two cores could potentially be harboring either very young or very low mass protostars (additional investigation is necessary to confirm 
the protostellar nature of MM05 and MM12). Our results thus suggest that these three cores are the least evolved in the Spokes cluster, i.e., closest to the very earliest star formation evolutionary phase. 
\cite{Friesen_2010-Deuterium_JCMT} found similar results in their analysis of the cores in the 
Ophiuchus B2 region, where the deuteration fraction decreases with proximity to protostellar cores. 
The spectrum of core MM03, which is located within the dense region of the Spokes cluster, shows two components, as mentioned.
Additional investigation, requiring higher angular and spectral resolution data, is needed to understand the 
deuteration fraction in this particular core.
Moreover, we found no correlation between the level of deuteration and velocity dispersion. 
Although these few data points make this result uncertain, they agree with what was found in 
\cite{Friesen_2013-Perseus_Deuteration} in a larger study of nearby low-mass cores.

\section{Summary}

We summarize our main results and conclusions as follows:
\begin{enumerate}

\item We observed in the Spokes cluster 14 cores in \nh(3--2) and 8 cores in \ndp(3--2)  with APEX; 
the measured linewidths are overall narrower than those of previous \nh(1--0) observations, 
$\sigma_{v}[\nh (3-2)] \sim 0.7\,\sigma_{v}[\nh(1-0)]$.

\item The three cores (MM03b, MM08, and MM11) that do not show narrower \nh(3--2) linewidths 
than the \nh(1--0) linewidths are cores with more than one component along the line-of-sight or low signal-to-noise spectra. 

\item The denser gas, probed by higher J-transitions of \nh and \ndp, presents lower levels of 
turbulence (nonthermal velocity dispersion) and in two cases approaches the sound speed. 
These results support the scenario of thermal fragmentation in the expectation that 
higher-angular resolution observations and/or higher-density tracers would find more quiescent gas.

\item Finally, we find that the three cores of our sample, spatially separated from the main Spokes cluster, are likely the youngest, based on their higher deuteration values.

\end{enumerate}

\begin{acknowledgements}
We thank Gary Fuller, Charlie Lada, Paola Caselli, and the anonymous referee for comments that improved the paper. 
We also thank the staff at the APEX telescope for performing the observations presented in this paper in service-mode.
JEP has received funding from the European CommunityÕs 
Seventh Framework Programme (/FP7/2007-2013/) under grant agreement No 229517. 
This publication is supported by the Austrian Science Fund (FWF).
\end{acknowledgements}

\appendix
\onecolumn
\section{Column density determination\label{sec-app-a}}
The column density is calculated for the optically thick and thin cases using different 
expressions, but in both cases assuming a beam filling factor of $1$. 
In the optically thick case, 
\begin{equation}
N_{Tot}=
2.01\times 10^{13} 
\left(\frac{\sigma_{v}}{\kms}\right)
\left(\frac{\mu}{\rm Debye}\right)^{-2}
\frac{1}{(J+1)}
\frac{\tau}{(1-e^{-T_{0}/T_{ex}})}\frac{Q_{rot}(T)}{e^{-J(J+1)h\,B/kT_{ex}}}
~, \label{eq-op-thick}
\end{equation}
and in the optically thin case, 
\begin{equation}
N_{Tot}=
8.01\times 10^{12} 
\left(\frac{W}{\rm K\, \kms}\right)
\left(\frac{\mu}{\rm Debye}\right)^{-2}
\frac{1}{(J+1)}
\left[\frac{1\,{\rm K}}{J_{\nu}(T_{ex})-J_{\nu}(T_{bg})}\right]
\frac{1}{(1-e^{-T_{0}/T_{ex}})}\frac{Q_{rot}(T)}{e^{-J(J+1)h\,B/kT_{ex}}}
~, \label{eq-op-thin}
\end{equation}
which are the same as eqs.  (A1) and (A4) in \cite{Caselli_2002-Ionization_L1544} after rearranging terms and using that 
\begin{eqnarray}
A_{J+1\rightarrow J}&=&\frac{64\pi^{4}}{3hc}\nu^{3}
|\mu_{J+1\rightarrow J}|^{2} = 
1.164 \times 10^{-11} \left(\frac{\nu}{\rm GHz}\right)^{3} \left(\frac{\mu}{\rm Debye}\right)^{2} \left(\frac{J+1}{2J+3}\right)~{\rm s^{-1}}\\
|\mu_{J+1\rightarrow J}|^{2} &=& \mu^{2} \frac{J+1}{2J+3}\\
g_{J}&=& 2J+1\\
E_{J}&=& J(J+1)hB\\
Q_{rot}(T)&=&\sum_{J=0}^{\infty}(2J+1)e^{-J(J+1)hB/kT}~,
\end{eqnarray}
where $T_{0}\equiv h\nu/k$ and $J_{\nu}(T)=T_{0}/[\exp(T_{0}/T)-1]$. 
The partition function is calculated using the first 100 levels.

\end{document}